\documentclass[12pt]{iopart}

\usepackage[dvips]{graphicx}
\DeclareGraphicsExtensions{.eps}
\usepackage{array}

\newcommand{\mus}{$\mathrm{\mu s}$}
\newcommand{\SciMode}{{\it science mode}}

\begin{document}
\title[]{Calibration and sensitivity of the Virgo detector \\
during its second science run}

\author{T.~Accadia$^{11}$, 
F.~Acernese$^{5ac}$, 
F.~Antonucci$^{8a}$, 
P.~Astone$^{8a}$, 
G.~Ballardin$^{2}$, 
F.~Barone$^{5ac}$, 
M.~Barsuglia$^{1}$, 
A.~Basti$^{7ab}$, 
Th.~S.~Bauer$^{13a}$, 
M.G.~Beker$^{13a}$
% ...
}
\address{$^{1}$Laboratoire AstroParticule et Cosmologie (APC) Universit\'e Paris Diderot, CNRS: IN2P3, CEA: DSM/IRFU, Observatoire de Paris 10, rue A.Domon et L.Duquet, 75013 Paris - France}
\address{$^{2}$European Gravitational Observatory (EGO), I-56021 Cascina (PI), Italy}
\address{$^{3}$INFN, Sezione di Firenze, I-50019 Sesto Fiorentino$^a$; Universit\`a degli Studi di Urbino 'Carlo Bo', I-61029 Urbino$^b$, Italy}
\address{$^{4}$INFN, Sezione di Genova;  I-16146  Genova, Italy}
\address{$^{5}$INFN, Sezione di Napoli $^a$; Universit\`a di Napoli 'Federico II'$^b$ Complesso Universitario di Monte S.Angelo, I-80126 Napoli; Universit\`a di Salerno, Fisciano, I-84084 Salerno$^c$, Italy}
\address{$^{6}$INFN, Sezione di Perugia$^a$; Universit\`a di Perugia$^b$, I-06123 Perugia,Italy}
\address{$^{7}$INFN, Sezione di Pisa$^a$; Universit\`a di Pisa$^b$; I-56127 Pisa; Universit\`a di Siena, I-53100 Siena$^c$, Italy}
\address{$^{8}$INFN, Sezione di Roma$^a$; Universit\`a 'La Sapienza'$^b$, I-00185 Roma, Italy}
\address{$^{9}$INFN, Sezione di Roma Tor Vergata$^a$; Universit\`a di Roma Tor Vergata, I-00133 Roma$^b$; Universit\`a dell'Aquila, I-67100 L'Aquila$^c$, Italy}
\address{$^{10}$LAL, Universit\'e Paris-Sud, IN2P3/CNRS, F-91898 Orsay$^a$; ESPCI, CNRS,  F-75005 Paris$^b$, France}
\address{$^{11}$Laboratoire d'Annecy-le-Vieux de Physique des Particules (LAPP), Universit\'e de Savoie, CNRS/IN2P3, F-74941 Annecy-Le-Vieux, France}
\address{$^{12}$Laboratoire des Mat\'eriaux Avanc\'es (LMA), IN2P3/CNRS, F-69622 Villeurbanne, Lyon, France}
\address{$^{13}$Nikhef, National Institute for Subatomic Physics, P.O. Box 41882, 1009 DB Amsterdam$^a$; VU University Amsterdam, De Boelelaan 1081, 1081 HV Amsterdam$^b$, The Netherlands}
\address{$^{14}$Universit\'e Nice-Sophia-Antipolis, CNRS, Observatoire de la C\^ote d'Azur, F-06304 Nice$^a$; Institut de Physique de Rennes, CNRS, Universit\'e de Rennes 1, 35042 Rennes$^b$, France}
\address{$^{15}$INFN, Gruppo Collegato di Trento$^a$ and Universit\`a di Trento$^b$,  I-38050 Povo, Trento, Italy;   INFN, Sezione di Padova$^c$ and Universit\`a di Padova$^d$, I-35131 Padova, Italy}
\address{$^{16}$IM-PAN 00-956 Warsaw$^a$; Warsaw University 00-681 Warsaw$^b$; Astronomical Observatory Warsaw University 00-478 Warsaw$^c$; CAMK-PAN 00-716 Warsaw$^d$; Bia{\l}ystok University 15-424 Bia{\l}ystok$^e$; IPJ 05-400 \'Swierk-Otwock$^f$; Institute of Astronomy 65-265 Zielona G\'ora$^g$,  Poland}
\address{$^{17}$RMKI, H-1121 Budapest, Konkoly Thege Mikl\'os \'ut 29-33, Hungary}
\address{$^{*}$University of Birmingham, Birmingham, B15 2TT, United Kingdom }
\address{$^{**}$University of Glasgow, Glasgow, G12 8QQ, United Kingdom }
\ead{rollandl@in2p3.fr}

\begin{abstract}
The Virgo detector is a kilometer-length interferometer for gravitational wave detection located near Pisa (Italy).
During its second science run (VSR2) in 2009, six months of data were accumulated with a sensitivity close to its design.

In this paper, the methods used to determine the parameters for
sensitivity estimation and gravitational wave reconstruction are described. 
The main quantities to be calibrated are the frequency response of the mirror actuation
and the sensing of the output power. Focus is also put on their absolute timing. 
The monitoring of the calibration data as well as the parameter estimation with independent techniques 
are discussed to provide an estimation of the calibration uncertainties.

Finally, the estimation of the Virgo sensitivity in the frequency-domain is described
and typical sensitivities measured during VSR2 are shown.
\end{abstract}

\pacs{95.30.Sf, 04.80.Nn}

\section{Introduction}
The Virgo detector~\cite{bib:VirgoDetector}, located near Pisa (Italy), is one of the most sensitive
instruments for direct detection of gravitational waves (GW) emitted by astrophysical 
compact sources at frequencies between 10~Hz and 10~kHz.
It is an interferometer (ITF) with 3~kilometers Fabry-Perot cavities in the arms.
Typical detectable length variations are of the order of $10^{-19}\,\mathrm{m}$.\\

The Virgo second science run (VSR2) lasted from July 3rd 2009 to January 8th 2010 with a
sensitivity close to its nominal one, 
and in coincidence to the first part of the 6th
science run (S6) of the LIGO detectors~\cite{bib:LigoDetectors}.
The data of all the detectors are used together to search for a GW signal. 
In case of a detection, the combined use of all the data
would increase the confidence of the detection 
and allow the estimation of the GW source direction and parameters.\\

The GW strain couples into the length degrees of freedom of the ITF. 
To achieve optimum sensitivity, the positions of the different mirrors 
are controlled~\cite{bib:LockAcquisition}  to have, in particular, 
beam resonance in the cavities, destructive interference at the ITF output port
and to compensate for environmental noise.
The control bandwidth would modify the ITF response to passing GW below a few hundreds hertz. 
Above a few hundred hertz, the mirrors behave as free falling masses ; 
the main effect of a passing GW would then be a frequency-dependent variation of the output power of the ITF,
characterized by the ITF optical response.\\

The main purposes of the Virgo calibration are
(i)   to estimate the ITF sensitivity to GW strain as function of frequency, $S_h(f)$,
(ii)  to reconstruct the amplitude $h(t)$ of the GW strain from the ITF data.
It deals with the {\it longitudinal}\footnote{
The ''longitudinal'' direction is perpendicular to the mirror surface.
} differential length of the ITF arms, $\Delta L=L_x-L_y$.
In the long wavelength approximation (see section 2.3 in~\cite{bib:Saulson1994}), it is related to the GW strain $h$ by
\begin{eqnarray}
h &=& \frac{\Delta L}{L} \quad\mathrm{where\,\,}L=3\,\mathrm{km}
\end{eqnarray}
The responses of the mirror actuation to the longitudinal controls have thus to be calibrated, 
as well as the readout electronics of the output power and the ITF optical response.
Absolute timing is also a critical parameter for multi-detector analysis, 
in particular to determine the direction of the GW source in the sky.\\

The scope of this paper is restricted to the estimation of the calibration parameters 
and to the description of the ITF sensitivity estimation in the frequency domain. 
The methods and performance of the calibration procedures are given after a brief description of the Virgo detector. 
The way the calibration parameters are used to estimate the Virgo sensitivity is then detailed. 
Finally, the Virgo sensitivity measured during VSR2 is presented.

\section{The Virgo detector}

\begin{figure}
\begin{center}
	\includegraphics[width=0.8\linewidth]{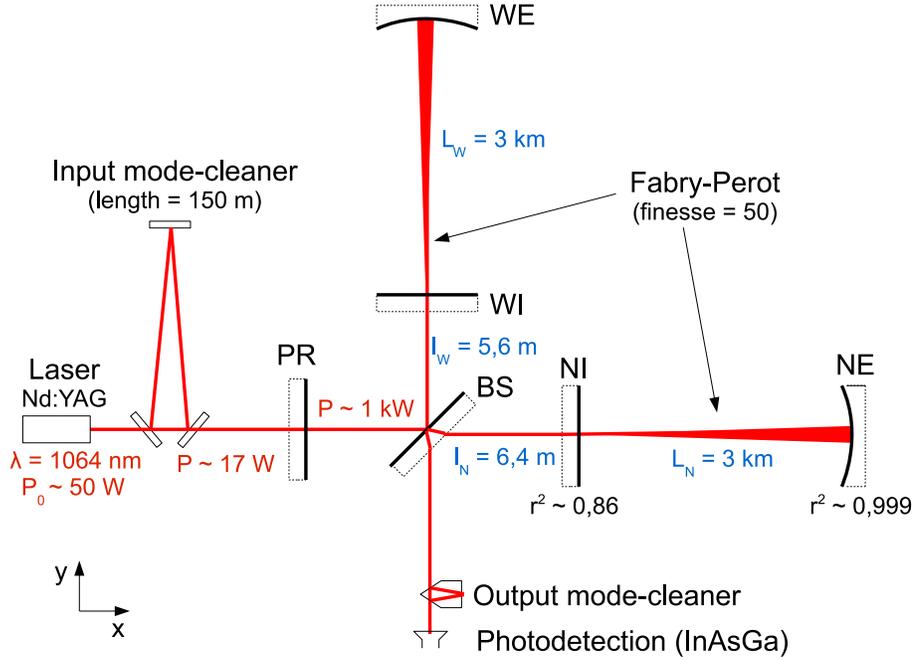}
	\caption{Optical scheme of Virgo.}
	\label{fig:OpticalScheme}
\end{center}
\end{figure}

The optical configuration of the Virgo ITF is shown figure~\ref{fig:OpticalScheme}.
All the mirrors of the ITF are suspended to a chain of pendulum for seismic isolation.
The input beam is produced by a Nd:YAG laser with a wavelength $\lambda=1064\,\mathrm{nm}$.
Each arm contains a 3-km-long Fabry-Perot cavity of finesse~50 whose role is to increase
the optical path.
The ITF arm length difference is controlled to obtain a destructive interference at the ITF output port.
The power recycling (PR) mirror increases the amount of light impinging on the Michelson beam splitter (BS) by a factor~40
~\cite{bib:VirgoDetector} and as a consequence improves the ITF sensitivity. 
The main signal of the ITF is the light power at the output port, the so-called {\it dark fringe} signal.

Data from the ITF are time series, digitized at 10~kHz or 20~kHz, 
which record the optical power of various beams and different control signals.

In order to analyse in coincidence the reconstructed GW-strain from different detectors,
the data are time-stamped using the Global Positioning System (GPS).

\subsection{Mirror longitudinal actuation}

The Virgo mirrors are suspended to a complex seismic isolation 
system~\cite{bib:SeismicAttenuation}. 
The last stage is a double-stage system with the so-called {\it marionette}~\cite{bib:LastStage}
as the first pendulum. The mirror and its reaction mass are suspended to the marionette by pairs
of thin steel wires.\\

Acting on the marionette, it is possible to translate the suspended mirror along the beam.
This steering is performed by a set of coils. 
Each coil acts on a permanent magnet attached on the marionette.

Four additional coils supported by the recoil mass allow to act 
directly on four magnets glued on the back of the mirror.
It induces a mirror motion keeping a fixed center of gravity of the suspended system \{mirror+recoil mass\}.\\

Above 10~Hz, the longitudinal controls are distributed between the marionette up to a few 10's of Hz
and the mirror up to a few 100's of Hz.
Therefore, for calibration purpose, the marionette and mirror actuation responses need to be measured only up to $\sim$100~Hz
and up to $\sim$1~kHz respectively.\\

The actuation is used to convert a digital control signal (called $zC$ hereafter, in V) into the mirror motion $\Delta L$
through an electromagnetic actuator and the pendulum.
The actuator is composed of a digital computing part (DSP), a DAC with its anti-image filter
and the analog electronics (so-called coil driver) which converts the DAC output voltage into a current
flowing in the coil. 
The electronics can be set into two configurations: 
(i)  a mode to acquire the lock of the ITF, the so-called high power (HP) mode ;
(ii) a low noise (LN) mode with reduced dynamic to control the lock of the ITF.
Different analog filters (poles and zeros) and resistors (gains) are used in the coil driver 
as function of the configuration.
Compensating digital filters and gains are set accordingly in the DSP such that the transfer function (TF)
of the actuator (in A/V) is independent of the configuration at first order\footnote{
Mis-compensations are measured, see section~\ref{lab:LNtoHP}}.\\

The mirror motion induced by the actuators is then filtered by the pendulum mechanical response.
The response to the mirror motion is modeled by a second order low-pass filter $P_{mir}$
with a resonant frequency 0.6~Hz and a quality factor arbitrarily set to 1000~\cite{bib:PendulumTF}. 
For the mirror motion induced through the marionette, 
the 2-stage pendulum $P_{mar}$ has been parameterized in the calibration procedure
as a series of two such filters.\\

The DC gain of the mirror actuation can be estimated within $\sim10\%$ from the nominal values 
of the conversion factors of the coil driver electronics $\gamma$, the current-force conversion factor $\alpha$
and the mechanical response of the pendulum. 
During VSR2, the conversion factors $\gamma$ and $\alpha$ are respectively
$1.15\,\mathrm{A/V}$ and $1.9 \,\mathrm{mN/A}$ for the end mirrors and
$0.10\,\mathrm{A/V}$ and $10.6\,\mathrm{mN/A}$ for the BS mirror~\cite{bib:NoteElectromagActuators}.
The pendulum can be modeled by a simple pendulum with length $l=0.7\,\mathrm{m}$ and with
mirror masses of $M_e = 20.3\,\mathrm{kg}$~\cite{bib:NoteVirgoSensitivity} 
and $M_{BS} = 5\,\mathrm{kg}$~\cite{bib:NoteBSPayload}  
for the end and BS mirrors respectively. 
The longitudinal motion of the mirrors is controlled by $n_{coils}$ coils: 
2 for the end mirrors and 4 for the BS mirror.
The estimated actuation gain, perpendicular to the mirror surface, is thus:
\begin{eqnarray}
A \,=\,\frac{F_V}{M \times \frac{g_N}{l}}\,=\, \frac{n_{coils} \gamma \alpha}{M \times \frac{g_N}{l}}
\end{eqnarray}
where $F_V$ the force per unit volt and $g_N=9.81\,\mathrm{m.s^{-2}}$ is the standard gravitational acceleration.
With the numbers given above, the gains for the end and BS mirrors are expected to be
$15.1\,\mathrm{\mu m/V}$ and $61\,\mathrm{\mu m/V}$ respectively.
For a motion $\Delta L_{BS}$ of the BS mirror, the length of the west arm is left unchanged while
the length of the north arm varies by $\sqrt{2}\,\Delta L_{BS}$. 
The effective gain on the differential arm length variation
is thus expected to be $61\,\sqrt{2} \sim 86\,\mathrm{\mu m/V}$.

\subsection{Sensing of the ITF output power}

In order to control the ITF~\cite{bib:LockAcquisition}, the laser beam is phase modulated. 
The main signal of the ITF is the demodulated output power, called $\mathcal{P}_{AC}$. 

The output power of the ITF is sensed using two photodiodes.
Their signals then go through pre-amplifiers, demodulation boards and ADCs with anti-alias filters. 
Both raw demodulated signals, digitized at 20~kHz, are then sent into a digital process 
where they are summed to compute the output port channel $\mathcal{P}_{AC}$.
In the following, the TF from the power at the output port to the measured signal
will be called~$S$. 

In the sensing process, the total power of the output beam is also readout and stored in the channel $\mathcal{P}_{DC}$.

\subsection{Longitudinal control loop}
\begin{figure}[tbh]
\begin{center}
	\includegraphics[width=0.8\linewidth]{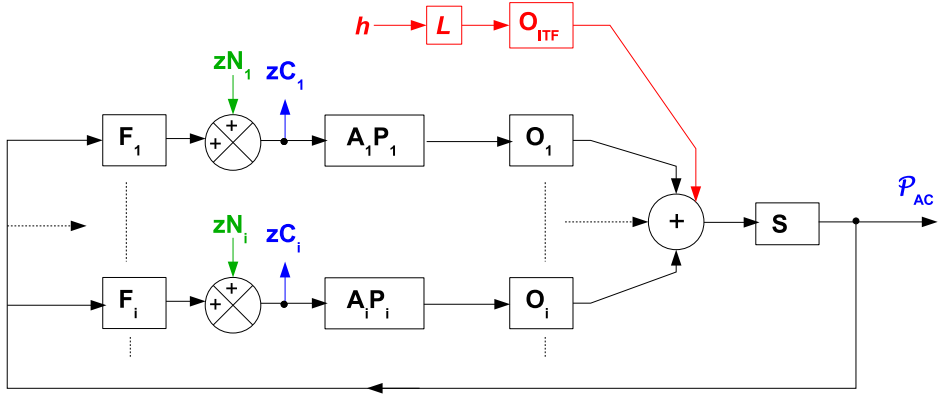}
	\caption{
	Overview diagram of the longitudinal control loop.
	For the actuation channel $i$: $A_i$ and $P_i$ are the actuator and pendulum responses,
	$O_i$ is the ITF optical response. $S$ is the TF of the sensing of the ITF output power $\mathcal{P}_{AC}$,
	used as error signal.
	$F_i$ is the TF of the global control loop.
	The actuation entries are the control signal and the calibration signal $zN_i$. 
	The sum of both gives the signal $zC_i$.
	The GW signal $h(t)$ enters the ITF as a differential motion of the two cavity end mirrors.
	}
	\label{fig:LongitudinalLoop}
\end{center}
\end{figure}

The longitudinal control loop~\cite{bib:LockAcquisition} used to lock the ITF on a {\it dark fringe} in \SciMode\ 
(standard data taking conditions) is summarized in figure~\ref{fig:LongitudinalLoop}.
The error signal is the ITF output power $\mathcal{P}_{AC}$ (W), readout through photodiodes and their associated
electronics with response $S$. Filters $F_i$ (V/W) are used to define the control signals sent to the different actuation channels
(controlled mirrors -NE, WE, BS, PR- and marionettes -NE, WE- indicated by the subscript $i$).
The signal is then sent to the actuator with response $A_i\times P_i$ (m/V) in order to move the mirror
($P_i$ is the mechanical response of the pendulum and $A_i$ the part of the response due to the electromagnetic actuator).
The ITF output power depends on the mirror positions and determines 
the optical response $O_i$ of the interferometer (W/m).

A calibration signal $zN_i$ can be added at the input of the actuation.
The sum of the control signal and of the calibration signal is readout as $zC_i$.

\section{Absolute length measurement}
\label{lab:freeMichTechnique}

\begin{figure}
\begin{center}
	\includegraphics[width=0.8\linewidth]{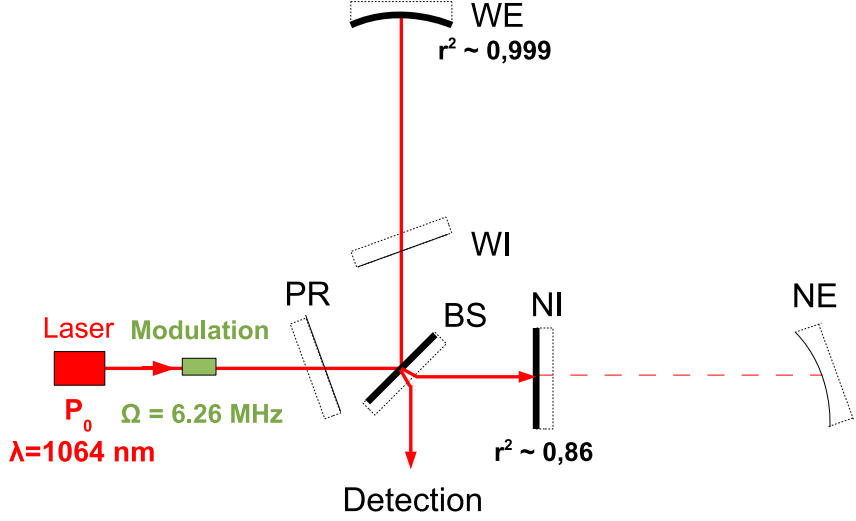}
	\caption{
	Example of asymmetric (NI-WE) Michelson configuration.
	The mirror reflective surfaces are shown with continuous lines.
	The surfaces that consitute the Michelson ITF are the bold lines.
	The tilted mirrors do not participate to the output signal (except for an attenuation of the laser power).
	}
	\label{fig:FreeMichConfig}
\end{center}
\end{figure}

The ITF calibration requires absolute length measurements.
The displacement induced by the mirror actuators is reconstructed from the ITF setup
as simple Michelson (see for example figure~\ref{fig:FreeMichConfig}) using the laser wavelength as length reference
and its non linear response when the fringes are passing.

The method is first described.
Corrections due to the ITF optical response and the power readout response are then highlighted.

\subsection{Output powers}

In a simple Michelson ITF, the phase difference $\Delta \Phi$ 
between the two interfering beams is a function of the differential arm length $\Delta L$,
and the laser wavelength $\lambda=1064\,\mathrm{nm}$:
\begin{eqnarray}
\label{eqn:FreeMichPhase}
\Delta \Phi(t) &=& \frac{4\pi}{\lambda} \Delta L(t) 
\end{eqnarray} \\

\begin{figure}
\begin{center}
	\includegraphics[width=0.6\linewidth]{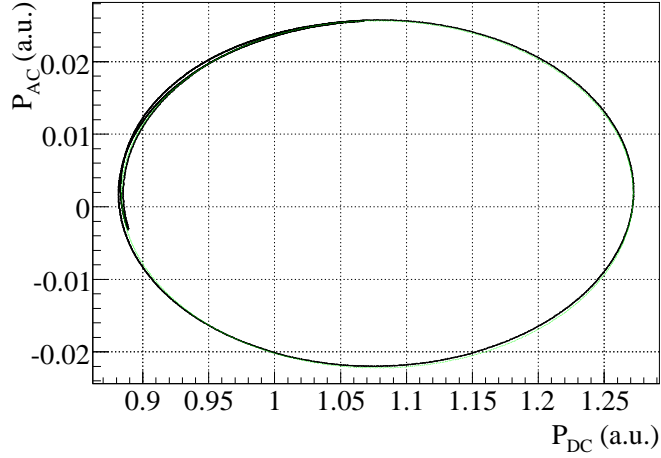}
	\caption{Example of the AC vs DC signals in a free swinging Michelson configuration
	(asymmetric Michelson) along with the fitted ellipse (green curve).	
	}
	\label{fig:FreeMichEllipse}
\end{center}
\end{figure}

In a simple Michelson ITF with frontal phase-modulation, 
the continuous ($\mathcal{P}_{DC}$) and demodulated ($\mathcal{P}_{AC}$) signals of the output beam
are function of the phase difference $\Delta\Phi$ between the two interfering beams~\cite{bib:freeMichNote}:
\begin{eqnarray}
\mathcal{P}_{DC}  &=& \beta(1-\gamma\cos(\Delta\Phi)) \\
\mathcal{P}_{AC}  &=& \alpha \sin(\Delta\Phi)
\end{eqnarray}
where $\alpha$ and $\beta$ are proportional to the laser power 
and $\gamma$ is proportional to the ITF contrast.
Therefore, in the ($\mathcal{P}_{DC},\mathcal{P}_{AC}$) plane, 
the signals follow an ellipse as shown in figure~\ref{fig:FreeMichEllipse}.

\subsection{Non-linear reconstruction method}

\begin{figure}[tbh]
\begin{center}
	\includegraphics[width=0.6\linewidth]{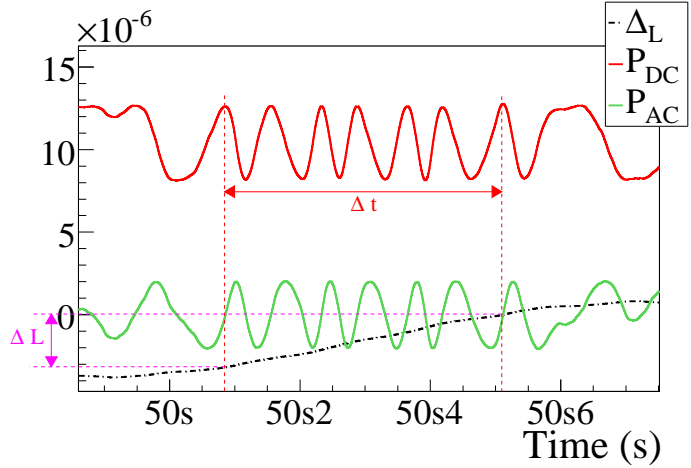}
	\caption{Reconstructed $\Delta L$ signal along with the DC and AC powers as function of time
	in asymmetric WE-NI Michelson configuration data.
	The unit of the y-axis depends on the signal: 1~m for $\Delta L$, 
	10~W for the AC power and 20~W for the DC power.
	In the window $\Delta t$, 6~interfringes passed on the DC signal.
	It indicates a differential arm elongation of $\Delta L = 6\times \frac{\lambda}{2} = 3.19\,\mathrm{\mu m}$.
	}
	\label{fig:FringeCounting}
\end{center}
\end{figure}

\begin{figure}[tbh]
\begin{center}
	\includegraphics[width=0.6\linewidth]{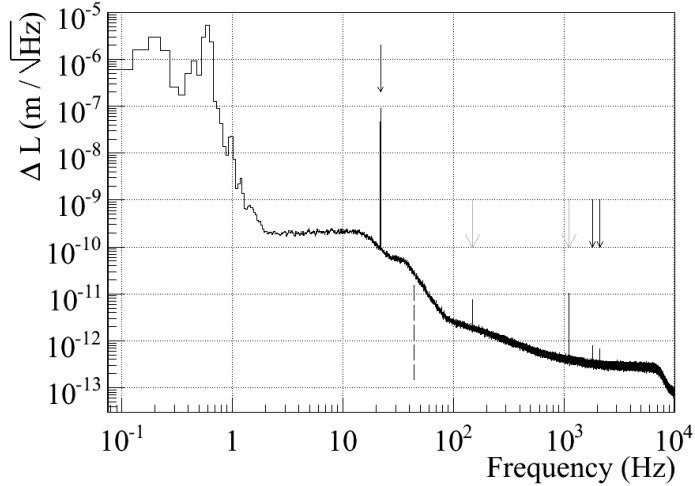}
	\caption{Typical noise level (FFT) of the reconstructed $\Delta L$ signal in free swinging Michelson data
	(asymmetric configuration).
	The resonance frequency of the pendulum is visible around $0.6$\,Hz.
	One power line is visible at 150~Hz.
	The line used for the laser frequency stabilization control loop is seen at 1111~Hz.
	The mirrors were excited at three frequencies shown by the dark arrows ($22.0$~Hz, $1816.5$~Hz and $2116.5$~Hz).
	In presence of non-linearities, the first harmonic of the strong 22~Hz line would have been visible at 44~Hz (dashed line).
	Note that the symmetric Michelson configuration using the two end mirrors, NE and WE, is not used since the noise level is higher.
	}
	\label{fig:EllipseMethodSensitivity}
\end{center}
\end{figure}

The measurement of the differential arm length $\Delta L$ uses a non-linear reconstruction.
The ellipse followed by the $\mathcal{P}_{AC}$ and $\mathcal{P}_{DC}$ signals is fitted~\cite{bib:HalirFlusser}. 
The fit gives the ellipse center position  and the axis widths.
In order to follow their possible time variations, the parameters are estimated every time the phase has changed by $2\pi$,
which correspond to a few times per second.
During the few minutes long datasets used for calibration, 
the typical variations are of the order of 1\% on the ellipse widths and center.
The angle between the ellipse axis along DC and the line from the ellipse center 
to the present point position ($\mathcal{P}_{DC},\mathcal{P}_{AC}$)
can then be estimated directly for every sample of the ITF signals.
Using a suitable ellipse tour counting, the right number of $2\pi$ is added
to recover completely the angle $\Delta\Phi$.
The differential arm length $\Delta L_{rec}$ is then computed using equation~\ref{eqn:FreeMichPhase}.\\
An illustration of the method is shown in figure~\ref{fig:FringeCounting}.\\

The typical sensitivity of the free swinging Michelson is given by the spectrum of the reconstructed $\Delta L_{rec}$ channel
as shown in figure~\ref{fig:EllipseMethodSensitivity}. 
The sensitivity is of the order of $10^{-9}$~m or below at few Hz and down to $3\times 10^{-13}$~m above 1~kHz.
The limiting noise sources have been determined from SIESTA simulation~\cite{bib:Siesta}. 
It comes from seismic noise below 1~Hz ; 
a mix of seismic noise and laser power noise up to $\sim$20~Hz ;
power noise up to $\sim$1~kHz ; and ADC noise at higher frequency.\\

The reconstruction method has been applied to simulated data produced by the SIESTA code.
No systematic error due to the method were found between 1~Hz and 10~kHz within 0.01\% 
when comparing the reconstructed $\Delta L$ with the simulated one.

Since the laser wavelength is precisely known, bias in the method could only arise between fringes.
When inducing sine motion to the mirrors, this would show up as harmonic lines in the spectrum of 
the reconstructed $\Delta L_{rec}$.
Such lines were not found above the noise level (see figure~\ref{fig:EllipseMethodSensitivity}). 
It indicates that possible non-linearities on the main line amplitude,
and therefore on the absolute length calibration, are lower than 0.1\%.\\

\subsection{Measured signals}
The differential mirror motion of the Michelson is converted by the ITF optical response $O_{Mich}$
into a power at the output port. The power is sensed with response $S$ to get the output channels $\mathcal{P}_{DC}$ and $\mathcal{P}_{AC}$.
Both responses have to be taken into account to reconstruct properly the differential length $\Delta L$:
\begin{itemize}
\item in free swinging Michelson configurations, $O_{Mich}$ is simply a delay due
to the light propagation time from the moving mirror to the photodiode: $10\,\mathrm{\mu s}$ for the end mirrors
(3~km) and 0 for the input mirrors (the propagation time in the central part can be neglected).
For the BS mirror, a delay of $10\,\mathrm{\mu s}$ is expected when the WE mirror is used, and no delay when the WI mirror is used.
\item below 2~kHz, $S$ is equivalent to a delay of $49.3\,\mathrm{\mu s}$ from the GPS time 
(see section~\ref{lab:OutputPortCalibration}).
\end{itemize}


\section{Actuation calibration}

The actuation response calibration consists in measuring the longitudinal mirror motion induced 
by an excitation signal sent to the mirror or marionette controls.
The actuation response, in m/V, can be written as the product of two TFs: $A_i \times P_i$ where
$P_i$ is the pendulum mechanical model.
For a better view of the measurements, only the part $A_i$ is shown in the following figures.
The full actuation TF, $A_i\times P_i$, permits to convert the input signal channel 
to the induced mirror motion with absolute timing using the GPS as reference.

When taking science data, the suspension electronics is set in LN mode.
The mirror actuation TF that is needed is thus the one using the LN mode electronics.
However, the range of the mirror actuation in LN mode is too low for direct measurements.
The measurements is thus done in different steps described in the following paragraphs: 
\begin{itemize}
\item the actuation TF is measured in HP mode from free swinging Michelson data,
\item the LN/HP actuation TF ratio is computed from measurements of the current flowing in the coils,
\item the actuation TF in LN mode is then computed combining these two measurements.
\end{itemize}

\subsection{Mirror actuation TF in HP mode}

\subsubsection{Procedure}

Data are taken in simple Michelson configurations using a single mirror in each arm:
the other arm mirror and the PR mirror are mis-aligned as shown in figure~\ref{fig:FreeMichConfig}.
The ITF is working in open-loop: no correction is applied to the mirrors nor marionettes.
Some sinusoidal signals (so-called {\it lines} it the following) $zN$ 
are applied to the mirror actuators set in HP mode.

The mirror motion $\Delta L_{rec}$ is reconstructed using the technique described in section~\ref{lab:freeMichTechnique},
taking into account the ITF optical response and the output power sensing.
Only good quality data with a coherence higher than 70\% between the excitation $zN$ and $\Delta L_{rec}$ are selected
to compute a first TF: $H=\Delta L_{rec}/zN$ in m/V.
Simulations were used to determine the errors of the TF modulus $M$ and phase $P$ (rad) from the coherence $C$ of the measurements as:
\begin{eqnarray}
\frac{\Delta M}{M} &=& a\sqrt{\frac{1-C}{C}}\frac{1}{\sqrt{n_{av}}} \\
\Delta P           &=& b\sqrt{\frac{1-C}{C}}\frac{1}{\sqrt{n_{av}}} 
\end{eqnarray}
where $n_{av}$ is the number of averages performed on the TF
and $a$ and $b$ two constants respectively estimated to~0.85 and~0.88.

\begin{figure}[tbh]
\begin{center}
	\includegraphics[width=0.6\linewidth]{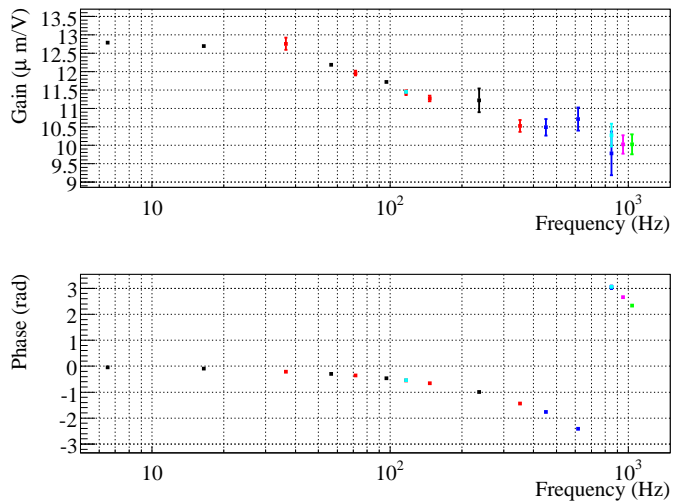}
	\caption{Typical mirror actuation TF: WE mirror using the Up-Down coils in HP mode, Sept. 1st, 2009
	  (the pendulum response is not included).
	\label{fig:WEActuationHP}
	}
\end{center}
\end{figure}

\begin{figure}[tbh]
\begin{center}
	\includegraphics[width=0.6\linewidth]{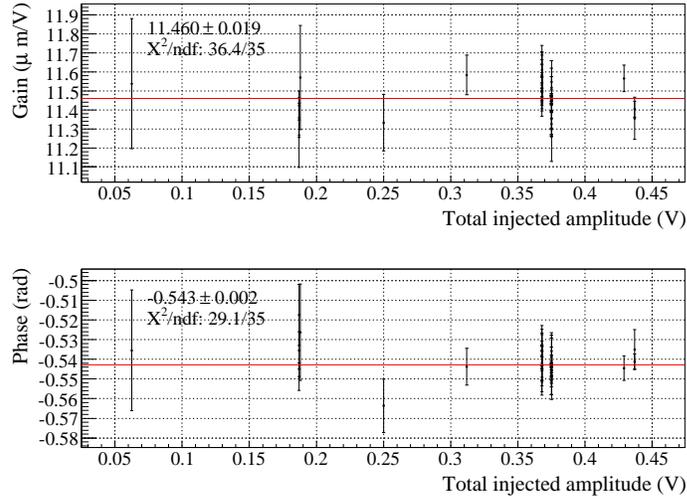}
	\caption{Linearity of the mirror actuation TF (WE mirror using the Up-Down coils in HP mode) 
	measured from June 2009 to January 2010: TF vs excitation amplitude (at 117~Hz).
	The average modulus and phase values are given and shown as the red line.
	}
	\label{fig:WEActuationHP_vsAmpl}	
\end{center}
\end{figure}

\begin{figure}[tbh]
\begin{center}
	\includegraphics[width=0.6\linewidth]{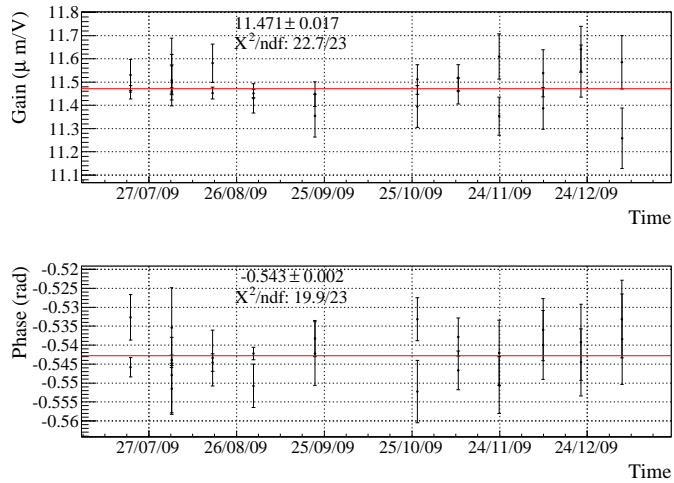}
	\caption{Stability of a mirror actuation TF (WE mirror using the Up-Down coils in HP mode) during VSR2, shown at 117~Hz.
	The average modulus and phase values are given and shown as the red line.}
	\label{fig:WEActuationHP_vsTime}
\end{center}
\end{figure}

\subsubsection{VSR2 measurements}

During VSR2, free swinging Michelson data have been taken every two weeks
to monitor the actuation TF, and a series of pre-run and post-run measurements was done in June 2009 and January 2010.
Lines were injected between 5~Hz and 2~kHz to measure the actuation response.
An example of mirror actuation TF measured during VSR2 is shown figure~\ref{fig:WEActuationHP}.\\

The linearity of the response was checked applying different amplitudes of the actuation excitation signal.
The evolution of the modulus of the actuation TF as function of the injected amplitude is 
shown figure~\ref{fig:WEActuationHP_vsAmpl}, along with the $\chi^2/ndof$ assuming a linear response.
It indicates that the response is linear within statistical errors.\\

The time stability of the actuation response during VSR2 has been monitored as shown in figure~\ref{fig:WEActuationHP_vsTime}.
In most cases, the $\chi^2/ndof$ assuming a constant response indicate that the modulus and phase 
are compatible with a constant during VSR2.
In few cases, time variations of the modulus were observed, but still below 1\%.
The actuation TF modulus and phase have thus been time-averaged.
Below 900~Hz, the statistical errors at the frequencies monitored during VSR2 are
below 1\% in modulus and 10~mrad in phase. 
Between 900~Hz and 2~kHz, they are estimated to 3\% and 30~mrad.

\subsection{LN/HP mirror actuation TF ratio}
\label{lab:LNtoHP}

\begin{figure}[tbh]
\begin{center}
	\includegraphics[width=0.6\linewidth]{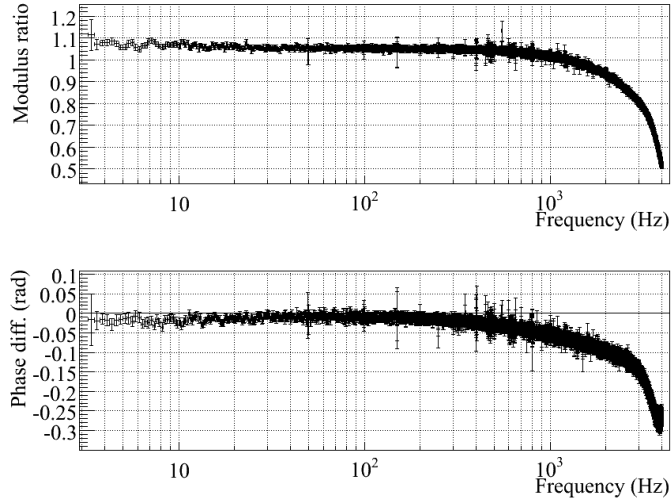} 
	\caption{LN to HP ratio measurement (WE mirror, coil Up, Sept. 1st, 2009).
	}
	\label{fig:WEActuation_LN1toHP}	
\end{center}
\end{figure}

\begin{figure}[tbh]
\begin{center}
	\includegraphics[width=0.6\linewidth]{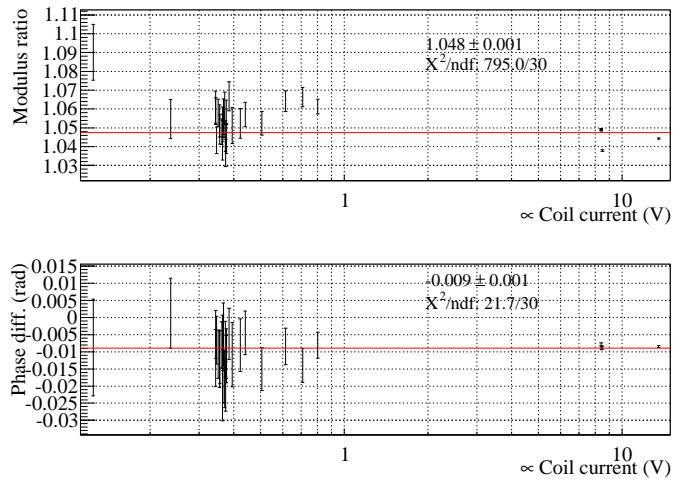}
	\caption{Linearity of LN to HP ratio (WE mirror, coil Up) 
	measured from June 2009 to January 2010: TF ratio vs excitation amplitude (at 117~Hz).
	The average modulus and phase values are given and shown as the red line.
	The $\chi^2/ndf$ assuming linearity is given.
	}
	\label{fig:WEActuation_LN1toHP_vsAmpl}
\end{center}
\end{figure}

\begin{figure}[tbh]
\begin{center}
	\includegraphics[width=0.6\linewidth]{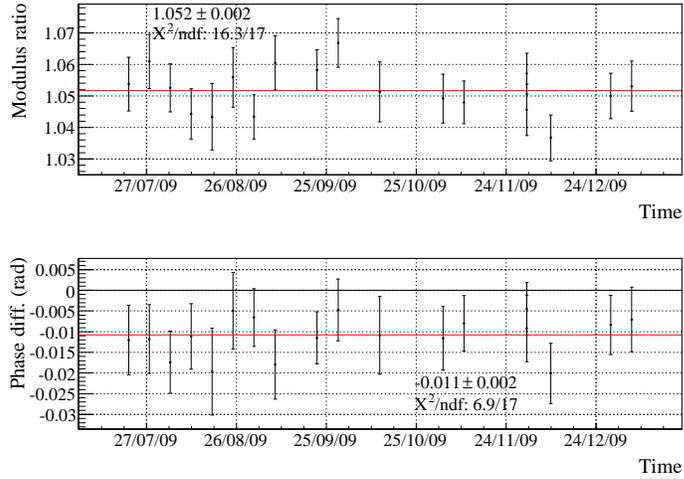}
	\caption{Stability of LN to HP ratio (WE mirror, coil Up) during VSR2, shown at  117~Hz.
	The average modulus and phase values are given and shown as the red line.
	The $\chi^2/ndf$ assuming constant response is given.
	}
	\label{fig:WEActuation_LN1toHP_vsTime}
\end{center}
\end{figure}

\subsubsection{Procedure}
The ratio of the LN to HP mirror actuator TF is necessary to convert the mirror actuation from the HP TF to the LN TF.
The differences of the actuation in both modes come from the different gains and filters in the DSP,
different DACs and different paths in the coil driver analog electronics.
The coupling of the coil-induced magnetic field with the mirror magnets and the pendulum response are not modified.
A direct way to measure the ratio is thus to inject an excitation $zN$ at the actuation input
(with no control signal),
and measure the current $C$ flowing in the coils in HP and in LN modes.

The ratio of the TFs $C_{LN}/zN$ to $C_{HP}/zN$ is a measurement of the actuator LN/HP ratio.
Since the current is measured through the same resistor and ADC whatever the mode is,
its sensing response cancels out in the ratio.
Such measurements were performed every week during VSR2.

\subsubsection{VSR2 measurements}
An example of LN/HP ratio measured during VSR2 is shown figure~\ref{fig:WEActuation_LN1toHP}.\\

Different amplitudes of the excitation were tested before and after VSR2 
in order to check the linearity and non-saturation of the electronics
as shown in figure~\ref{fig:WEActuation_LN1toHP_vsAmpl}.
Only three coil actuations out of twelve present possible non-linearities which are nevertheless
lower than 1.5\% in modulus and 10\,\mus\ in phase\footnote{The phase non-linearities are frequency dependent
and have been estimated as a delay $t_d = \phi / (2\pi f)$}.
These numbers are conservatively used as systematic errors for all mirror actuations.\\

The stability of the LN/HP ratio was monitored during VSR2 as shown in figure~\ref{fig:WEActuation_LN1toHP_vsTime}.
Since no time variation was found, the measurements have been time-averaged. 
The statistical errors are below 1\% in modulus and 10\,mrad in phase up to a few kHz.

\subsection{Mirror actuation TF in LN mode}
\label{lab:MirrorActuation_LNMode}

\begin{figure}[tbh]
\begin{center}
	\includegraphics[width=0.8\linewidth]{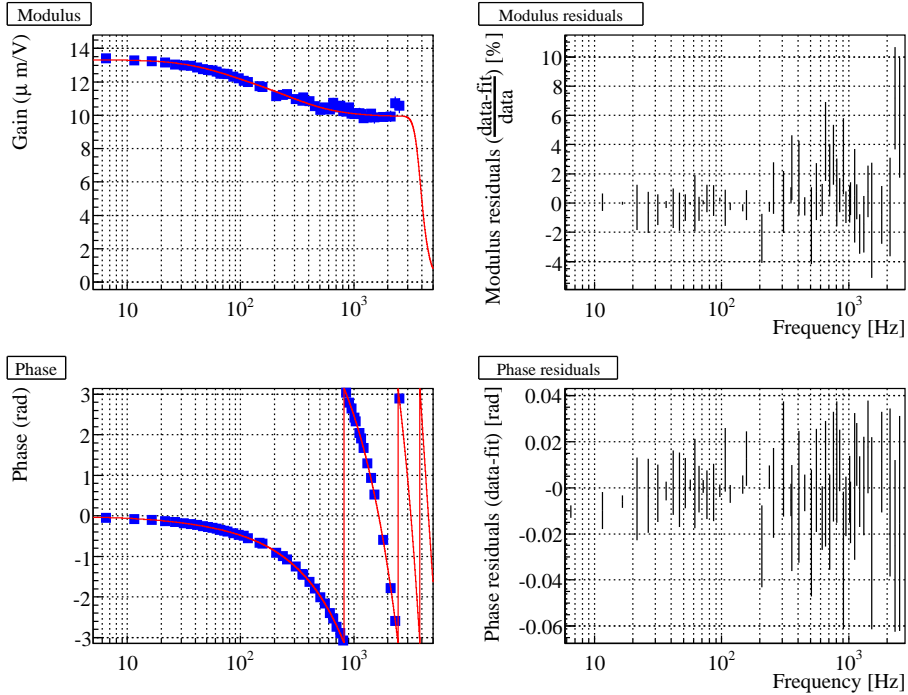}
	\caption{WE mirror actuation in LN mode.
		On the left column, the modulus and phase measurements are given (blue squares)
		along with the fitted parameterization (red curve).
		On the right column, the modulus and phase residuals of the parameterization are shown.
		The fit results in a DC gain of $13.33\pm0.02\,\mathrm{\mu m/V}$,
		two simple poles and two simple zeros around 70~Hz, 80~Hz and 300~Hz, 350~Hz respectively,
		and a delay of $271.3\pm0.7\,\mathrm{\mu s}$.
		}
	\label{fig:WEActuationLN}
\end{center}
\end{figure}

The mirror actuation TF in LN mode can be derived multiplying the averaged actuation TF in HP mode and the averaged LN/HP TF ratio.
Only the statistical errors are used.
The modulus and phase of the obtained TF are then fit simultaneously with a parameterization including
the nominal DAC anti-imaging filter (6th order elliptical filter with a cut-off at 3.7~kHz)
as well as free parameters: a gain, a delay, and simple poles and simple zeros. 
Poles and zeros are arbitrarily added such that the fit matches the data with a high $\chi^{2}$ probability:
the residuals are thus within the statistical errors of the order of 1.4\%/14\,mrad in modulus and phase below 900~kHz
and of 3.2\%/32\,mrad up to 2~kHz.

The example of the WE mirror actuation TF data, fit and residuals in LN mode is shown in figure~\ref{fig:WEActuationLN}.
The fit results in a DC gain of $13.33\pm0.02\,\mathrm{\mu m/V}$, in agreement with the expected value.
The fit $\chi^2$ is 48.0 for 88 degrees of freedom. It indicates that the parameterization is compatible with the data.
Similar results have been obtained on the other mirror actuations~\cite{bib:NoteCalibVSR2_Actuation}.
Therefore no additional systematic error on the actuation in LN mode is estimated from the fit residuals.\\

\begin{figure}[tbh]
\begin{center}
	\includegraphics[width=0.6\linewidth]{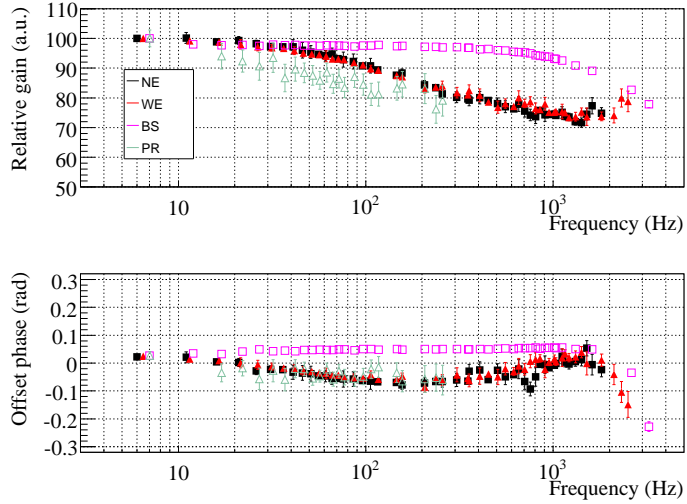}
	\caption{Measured mirror actuation in LN mode.
		The modulus have been normalised to the fitted DC gain.
		For better visibility, a delay of  $(270+180)\,\mathrm{\mu s}$ has been subtracted 
		from the phase
		($\sim180$\,\mus\ is introduced by the DAC anti-imaging below 1~kHz. 
		$270$\,\mus\ is close to the delays fitted on the measurements).
		The statistical errors are shown.
		WE: black squares.
		NE: red full triangles.
		BS: pink empty squares.
		PR: green empty triangles.
		}
	\label{fig:ActuationLN_All}
\end{center}
\end{figure}

The measured actuation TFs are all superposed in figure~\ref{fig:ActuationLN_All}.
The shape of the TF modulus was expected to be flat from the electronics point of view.
However, a frequency dependence has been observed, which is different for the BS actuation compared to the other mirror actuation. 
It is explained by the presence of eddy currents induced in the reference mass (RM) by the coil current. 
The shape and order of magnitude of the effect is in agreement with a finite element simulations of the system.
The effect is expected to be lower for the BS mirror since the RM is made of Aluminium
while the arm RMs are made of stainless steel. 
The geometry of the coil supports also induces lower eddy currents in the BS RM.
The data confirm this qualitative expectation.

In the 1~kHz--2~kHz region, the modulus of the mirror actuation responses tend to increase by a few percents. 
This might be related to the excitation of internal modes of the mirrors~\cite{bib:NoteFEMmirrors}.
For the end mirrors, the resonance frequencies are close to 5.5~kHz for the drumhead mode
and to 3.9~kHz for the butterfly mode.

\subsubsection{Estimation of systematic errors}

\begin{figure}
\begin{center}
	\includegraphics[width=0.6\linewidth]{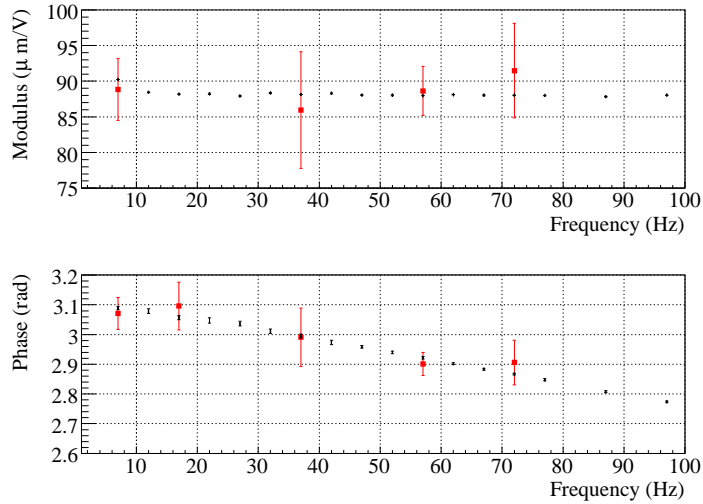}
	\caption{Comparison of the BS mirror actuation response measured with
	(black) the standard method and
	(red squares) direct measurements with free swinging Michelson data.
	}
	\label{fig:ActuationCheck_BSMirror}
\end{center}
\end{figure}

\paragraph{Free swinging Michelson in LN mode - }
Direct measurements of the mirror actuation TF in LN mode have been performed below 100~Hz using free swinging Michelson data.
Due to low mirror excitation level in LN mode, it was only possible to get measurements for the BS mirror. 
The comparison with the LN actuation TF of the BS mirror obtained
by the main method is shown in figure~\ref{fig:ActuationCheck_BSMirror}.
It did not indicate any systematic difference within the statistical
errors of the order of 5\% on the modulus and 50~mrad on the phase.

\paragraph{Estimated errors - }
The statistical and systematic errors obtained on the BS and end mirrors actuation TF measurements
during VSR2 are summarized in table~\ref{tab:ActuationErrors}.
\begin{table}[hbp]
\begin{center}
\caption{Statistical and systematic errors of mirror actuation TF measurements.}
\begin{tabular}{|c|c|c|c|c|}
\hline
               &  \multicolumn{2}{ c|}{5\,Hz$-$900\,Hz}   &  \multicolumn{2}{c|}{900\,Hz$-$2\,kHz}  \\
               &  stat.      & syst.                      &  stat.      & syst.                       \\     
\hline
Modulus        & 1.4\%       & 2.5\%                      & 3.1\%       & 2.5\%                       \\
Phase          & 14\,mrad    & 10\,\mus\                  & 31\,mrad    & 10\,\mus\                   \\
\hline
\end{tabular}
\label{tab:ActuationErrors}
\end{center}
\end{table}

The statistical error is quadratically summed to the systematic error to estimate 
the error on the mirror actuation modulus to $3\%$ below 900~Hz and $4\%$ above.
The phase error is dominated by a constant phase of 14~mrad below $\sim200$~Hz and by a delay of 10\,\mus\ above.

\subsection{PR mirror actuation TF}
Since the PR mirror cannot be calibrated from a free swinging Michelson configuration, an indirect method has been used. 

First, the ratio of the PR actuation TF to the BS actuation TF is extracted locking the PR-WI cavity:
the BS is used as a folding mirror. A motion of PR, BS or WI has the effect of changing the
length of the cavity. This can be sensed by the photodiode at the ITF output, which is used
for the locking of the cavity. 
Adding the same signal to the corrections sent to BS or PR has the same effect in terms
of changing the cavity length (i.e. the optical response), provided a factor $\sqrt{2}$ is taken into account because
of the $45^{\circ}$ orientation of the BS mirror.

Two datasets were taken at the end of VSR2. 
One with excitation injected on the PR mirror actuation and one with excitation injected on the BS mirror actuation.
The TFs of the excitation $zN$ to the ITF output $\mathcal{P}_{AC}$ can be written:
\begin{eqnarray*}
TF\bigg[\frac{\mathcal{P}_{AC}}{zN_{PR}} \bigg] &=& \frac{S O P_{PR} A_{PR}}{1-G_{olg}} \\
TF\bigg[\frac{\mathcal{P}_{AC}}{zN_{BS}} \bigg] &=& \frac{S O P_{BS} A_{BS}}{1-G_{olg}}
\end{eqnarray*}
with the open-loop gain $G_{olg}=S O \sum_i P_i A_i F_i$ where $i$ stands for the three mirrors of the cavity. 
The sensing response $S$ and the optical response $O$
are the same for all the mirrors as well as the pendulum mechanical response $P_i$.
The ratio of both TFs thus gives a measurement of $ A_{PR}/A_{BS}$.
The ratio has been measured below 300~Hz with statistical errors of the order of 5\%/50~mrad.\\ 

As as second step, the ratio is multiplied by the BS actuation TF to get the PR actuation TF.
The shape of the PR mirror actuation is shown in figure~\ref{fig:ActuationLN_All}.
Below 300~Hz it is compatible with the arm mirror actuation TFs,
as expected since they have the same design.

\subsection{Marionette actuation TF}

\begin{figure}[tbp]
\centering
	\includegraphics[width=0.6\linewidth]{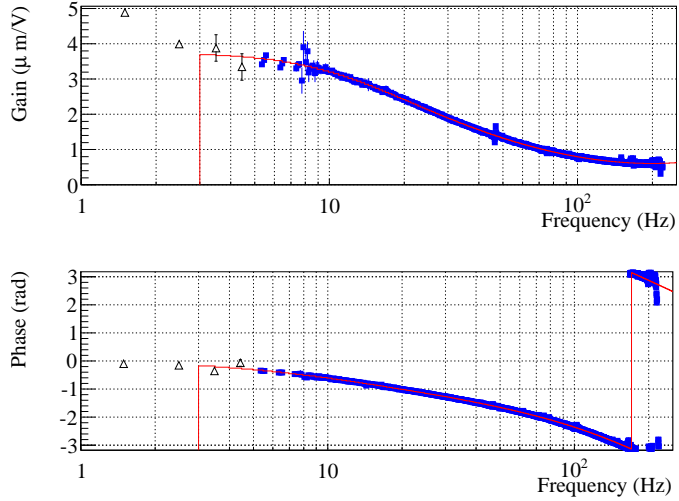}
        \caption{WE marionette actuation TF.
	Blue squares: measured modulus and phase.
	Red curve: the fitted model (shown for $f>3$~Hz).
	Empty triangles: direct measurements using free swinging Michelson data are also shown for comparison.
        }
	\label{fig:MarionetteFit}
\end{figure}

\subsubsection{Procedure}
Due to the low-pass mechanical response of the double-stage pendulum,
the marionette actuation TF cannot be measured directly using free swinging Michelson data.
The actuation TF is thus measured in two steps.
Once the mirror actuation TF is computed, the ratio of the marionette actuation TF
to the mirror actuation TF is measured. The marionette actuation is then derived.\\

Excitation has been injected on a mirror and then on its marionette
while the ITF is locked as in \SciMode. 
Using the description from figure~\ref{fig:LongitudinalLoop}, the actuation channels are respectively 
called $k_{mir}$ and $k_{mar}$ in the following.
The TFs $T_{k_{mir}}$ and $T_{k_{mar}}$ from the excitation $zN$ to the ITF output can be written:
\begin{eqnarray*}
T_{k_{mir}}\,=\,TF\bigg[\frac{\mathcal{P}_{AC}}{zN_{k_{mir}}}\bigg] &=& \frac{S O_{k} P_{k_{mir}} A_{k_{mir}}} {1-G_{olg}} \\
T_{k_{mar}}\,=\,TF\bigg[\frac{\mathcal{P}_{AC}}{zN_{k_{mar}}}\bigg] &=& \frac{S O_{k} P_{k_{mar}} A_{k_{mar}}} {1-G_{olg}} 
\end{eqnarray*}
with the open-loop gain $G_{olg}=S\sum_i O_i P_i A_i F_i$. Since the excitation is injected on the
mirror and its marionette, the ITF optical response is the same for both.
The actuation TF ratio, corrected for the different responses of the pendulum to the mirror and marionette motions,
is then derived as:
\begin{eqnarray*}
\frac{A_{k_{mar}}} {A_{k_{mir}}} &=& \frac{P_{k_{mir}}}{P_{k_{mar}}} \frac{T_{k_{mar}}}{T_{k_{mir}}}
\end{eqnarray*}

The marionette actuation TF $A_{k_{mar}}$, in m/V, is then computed multiplying this ratio by the 
parameterization of the corresponding mirror actuation TF $A_{k_{mir}}$.

\subsubsection{VSR2 measurements}
The controls of the end mirrors through the marionettes are negligible above a few 10's of Hz.
During VSR2, they have thus been measured from 5 to 200~Hz.

Different amplitudes of the excitation signal were tested.
No non-linearities in the ratio of the marionette to mirror actuation responses were observed within statistical errors.
Since no time variations were observed in the weekly monitoring during VSR2, the measurements have been averaged.
The statistical errors of the ratio are of the order of 3\% in modulus and 30\,mrad in phase.

The example of the WE marionette actuation TF and fit is given in figure~\ref{fig:MarionetteFit}.
The TF has been fit from 5~Hz to 150~Hz, with free parameters: a gain, a delay and simple and complex poles and zeros. 
The fit residuals are within statistical errors up to 100~Hz.
Similar results were obtained on the NE marionette~\cite{bib:NoteCalibVSR2_Actuation}.

\subsubsection{Marionette actuation systematic errors}
Since the parameterization of the mirror actuation TF is used to measure the marionette actuation TF,
the systematic errors are at least the errors from this parameterization: 3\% and 14~mrad on the modulus and the phase.
Additional source of errors were searched comparing the results with other measurements.

\paragraph{Free swinging Michelson measurements -}
Direct measurements of the marionette actuation TF have been performed injecting lines to the marionette
in free swinging Michelson configurations. Due to the low-pass mechanical response of the double-stage pendulum,
the measurements were possible from 1.5~Hz to 5~Hz only. 
The measurements are compared to the standard ones in figure~\ref{fig:MarionetteFit}.
Since there is no frequency overlap, only a rough check that the modulus and phase do not show any offset between
both measurements could be done, within errors of the order of 10\% and 100~mrad respectively.


\section{Output port calibration}
\label{lab:OutputPortCalibration}

The main ITF signal, $\mathcal{P}_{AC}$, is a measurement of the power at the ITF output port ({\it dark fringe} signal).
In this section, the stability of the Virgo timing system, which is critical for the global GW search, is first studied.
The calibration of the output power sensing and its absolute timing is then described.

\subsection{Virgo timing system}
èè
The Virgo data acquisition system~\cite{bib:DAQpaper} (DAQ) 
and timing system installed before VSR2 have been described in~\cite{bib:NoteVirgoPTiming}.
The timing system is based on a master timing system controlled by GPS.
Its roles are to give the rythm of the control loops and to give the time stamps to the DAQ.\\
The GPS receiver delivers a 1 pulse-per-second (PPS) clock and the corresponding date encoded in the IRIG-B format~\cite{bib:IRIG-B}.
This signal is distributed to all active elements (i.e. ADC, DAC) located in the four ITF buildings
with the same propagation delay, measured~\cite{bib:NoteVirgoPTiming} to $16.041\,\mathrm{\mu s}$.

\subsubsection{Stability of the timing system}
\begin{figure}[tbp]
\centering
	\includegraphics[width=0.6\linewidth]{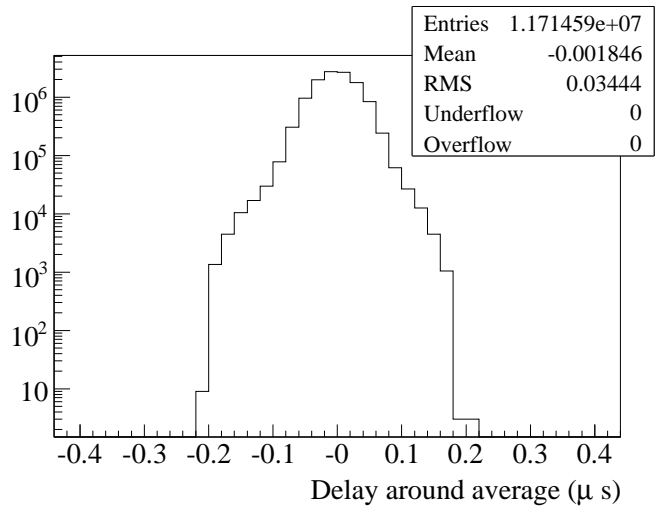}
        \caption{Fluctuations of the delays measured between the Virgo timing system 
	and the drift-corrected atomic clock. 
	Data during \SciMode\ segments of VSR2 have been used.
        }
	\label{fig:TimingStability_Atomic}
\end{figure}

In order to monitor the stability of the timing system over periods longer than 1~s, the 1~PPS signal delivered by 
an independent atomic clock is used.
The delay between the start of the 1~PPS signal and the start of the new second in the Virgo data is 
measured~\cite{bib:NoteTimingStabilityVSR2}.\\
The monitoring atomic clock being independent, its clock slowly drifts from the GPS reference.
The drift is estimated assuming it is linear over periods of one day.
The distribution of the delays around the average drift monitored during the full VSR2 
is shown in figure~\ref{fig:TimingStability_Atomic}.
No tails are observed outside $\pm0.3$\,\mus, which is consistent with the precision of the
IRIG-B signal, of 100~ns.\\   

As an other monitoring of the timing system, the corrections received every second by the ADC
boards to re-synchronize their local clock to the main 1~PPS clock have been checked~\cite{bib:NoteTimingStabilityVSR2}.
No timing instabilities have been found during VSR2, within again a precision of $0.3$\,\mus.\\

Over periods shorter that 1~s, the timing jitter of the main 1~PPS clock has also been measured~\cite{bib:NoteCalibVSR2_Sept09}
to be below $0.1\,\mathrm{ns/\sqrt{Hz}}$, orders of magnitude below the calibration timing uncertainties.

\subsection{Sensing electronics}
\label{lab:ReadoutElectronicsTF}

\begin{figure}
\begin{center}
	\includegraphics[width=0.8\linewidth]{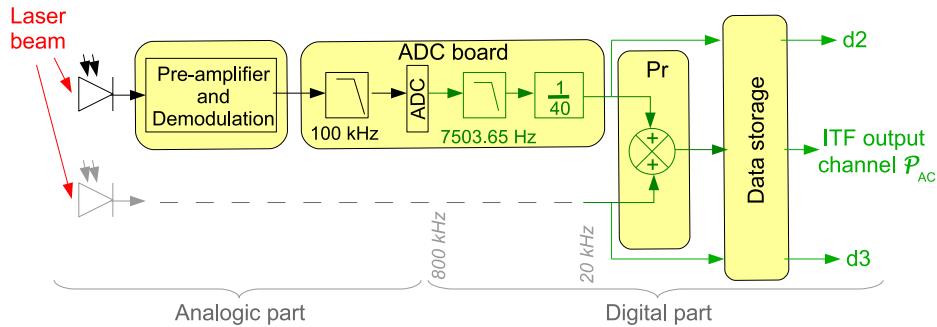}
	\caption{
	Overview diagram of the sensing of the ITF output power sampled at 20~kHz.
	}
	\label{fig:PhotodiodeSensing}
\end{center}
\end{figure}

The sensing scheme of the ITF output power is shown in figure~\ref{fig:PhotodiodeSensing}.
Its TF must be taken into account in the GW strain reconstruction in order not to distort 
the reconstructed amplitude and phase.
The readout is composed of:
\begin{itemize}
\item the photodiodes and the analog electronics.
\item the ADC-board with
(i)   an analog anti-alias filter (6th order Butterworth filter) with cut-off frequency at 100~kHz,
(ii)  the ADC sampling the signal at 800~kHz,
(iii) a digital 8th order Butterworth filter with cut-off frequency at 7503.65~Hz,
(iv)  a digital decimation of the signal at 20~kHz picking 1 sample over 40 (the last sample of the 40 of the window is kept)
and (v)  time stamping.
\item the DAQ to store the data. It does not introduce any delay.
\end{itemize}

The response of the photodiode and analog electronics before the ADC board is assumed to be flat in the frequency band of Virgo
and to introduce negligible delays.
The response of the ADC-board analog anti-alias filter was measured to be equivalent to a delay of 
$5.7\pm0.2$\,\mus\ below 10~kHz. 
The response of the digital processes has been precisely checked and can be perfectly modeled up to 10~kHz
by the digital Butterworth filter and an advance of $48.75\,$\mus, introduced by the digital decimation.
Below 2~kHz, the Butterworth filter can be approximated by a delay of $109\pm1$\,\mus.

\subsubsection{Timing of the readout electronics}
\label{lab:ReadoutElectronicsTiming}
The knowledge of the absolute delay introduced by the sensing of the output port laser power is essential
in order to provide absolute timing of the Virgo data.

From our understanding of the sensing described above, the sensing below 2~kHz can be modeled by a
delay of $5.7+109-48.75=66.0$\,\mus. 
In order to take the GPS time as reference, an advance of 16\,\mus\ must be added due to the timing distribution system:
it results in a total delay of $50.0$\,\mus.

In order to measure this delay, the 1~PPS clock from the main GPS receiver of the Virgo timing system
has been reshaped to a ramp and digitized by the ADC board also used to sample the output power $\mathcal{P}_{AC}$.
The beginning of the ramp, estimated from a fit, indicates the time of the 1~PPS.
The delay between the start of the second in the Virgo data and the 1~PPS is a measurement of the readout delay.
The delay introduced by the analog part of the ADC board has been 
measured using the raw 800~kHz ADC values~\cite{bib:NoteCalibVSR2_Sept09}.
The measurement is in agreement with the expected value, within the $\pm4$\,\mus\ systematic errors introduced by the measurement method.
It has been checked that the digital processing of the 800~kHz values behaves as expected and do not introduce further timing uncertainties.

During VSR2, the 1~PPS ramped signal has been continuously sampled at 20~kHz~\cite{bib:NoteTimingStabilityVSR2}. 
The distribution of the measured delays
around its average value has a RMS of less than 40~ns and tails within 0.3\,\mus,
again compatible with the precision of the IRIG-B signal.

The photodiode power readout is thus known within $\pm4$\,\mus\ and was stable within 300~ns during VSR2.

\subsubsection{Models for ITF sensing}
Following the studies described in the previous sections, 
the ITF photodiode sensing below 2~kHz can be parameterized by a simple delay of $49.3$\,\mus,
taking as reference the GPS time.
The full model, valid up to 10~kHz, is a delay of $-59.7$\,\mus\ and a 8th order Butterworth filter
with cut-off frequency at 7503.65~Hz.
Systematic errors on the absolute timing are estimated to $4\,$\mus.


\section{Estimation of the sensitivity during VSR2}

One of the final results of the calibration is the estimation of the detector noise level as a function of frequency,
the so-called sensitivity curve. It corresponds to the noise level of the detector in terms of GW strain signal 
in the frequency domain, $\tilde{h}(f)$.
The sensitivity is estimated directly in the frequency-domain, without reconstructing the $h(t)$ signal
in the time-domain. 
In this section, measurement of the global response of the ITF to a GW strain is first presented 
and then used to estimate the sensitivity curve. 
The behaviour of the Virgo sensitivity during VSR2 is finally described.\\

\subsection{Virgo transfer function}
\label{lab:VirgoTF}

\begin{figure}[tbh]
\begin{center}
	\includegraphics[width=0.6\linewidth]{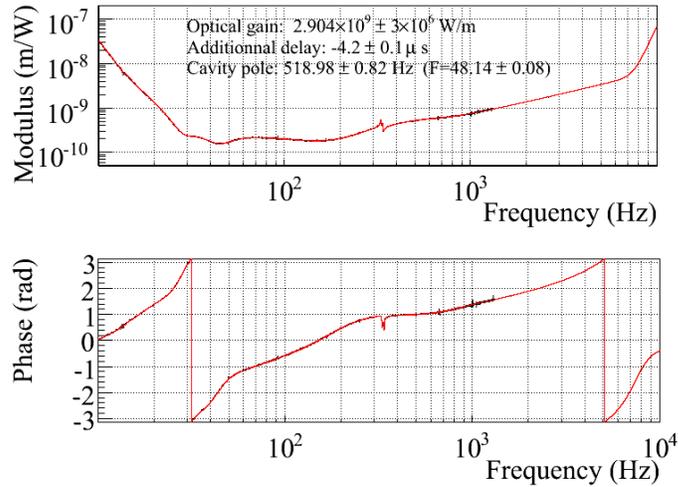}
	\caption{Virgo TF: modulus and phase
	in the range 10~Hz to 10~kHz (December 29th 2009).
	The data are in black (only the points with coherence higher than 70\% are shown). 
	The continuous red line is the Virgo TF model: it is a copy of the measurements below 900~Hz, 
	and is then extrapolated at higher frequency.
	}
	\label{fig:VirgoTF}
\end{center}
\end{figure}

The global TF of the ITF describes the path from the differential arm elongation $\Delta L$ in meter 
to the output power channel $\mathcal{P}_{AC}$. It is measured from specific data taken with the detector
in the same configuration as in \SciMode. Excitation (colored noise) is injected to the actuation of
one end mirror (for example the mirror $j$), through the channel $zN_j$
(see figure~\ref{fig:LongitudinalLoop}). 
The inverse of the ITF TF is defined, in W/m, as:
\begin{eqnarray*}
TF\bigg[\frac{\mathcal{P}_{AC}}{\Delta_L} \bigg] &=& TF\bigg[\frac{\mathcal{P}_{AC}}{zN_j}\bigg] \frac{1}{A_j P_j}
\end{eqnarray*}

The TF $\frac{\mathcal{P}_{AC}}{zN_j}$ is measured directly from the data, from a few Hz to $\sim1.5\,\mathrm{kHz}$.
The mirror actuation response $A_j$ (in LN mode) and the pendulum mechanical response $P_j$ are known
from the actuator calibration (see section~\ref{lab:MirrorActuation_LNMode}). \\

Above 1.5~kHz, the TF has to be extrapolated. It can be described in more details as:
\begin{eqnarray*}
TF\bigg[\frac{\mathcal{P}_{AC}}{\Delta_L} \bigg] &=& \frac{S O_j P_j A_j}{1-G_{olg}}\times\frac{1}{A_j P_j}
\end{eqnarray*}
In this frequency band, the longitudinal motion of the ITF mirrors is free ($G_{olg}\sim0$):
at 1~kHz, the contribution from the controls to the ITF differential motion has been estimated to be of the order of 1\%.
The ITF TF can thus be simply described by $S\times O_j$, the combination of the photodiode readout response and
of the ITF optical response to the mirror motion.
The response of the photodiode sensing has been precisely calibrated (see section~\ref{lab:ReadoutElectronicsTF}).

The optical response of the ITF to a mirror displacement~\cite{bib:OpticalResponseModel} 
is approximated by a simple pole whose frequency depends
on the average cavity finesse following:
\begin{eqnarray*}
f_p &=& \frac{c}{4 L F}
\label{eqn:PoleVsFinesse}
\end{eqnarray*}
where $c$ is the light speed and $L$ the cavity length (3~km).
For a finesse of~50, the cavity pole is close to 500~Hz but is expected to vary by $\sim\pm3\%$ 
due to etalon effect in the input mirror varying with the mirror temperature
(the input mirrors are flat-flat mirrors with non-zero reflectivity on the face with anti-reflecting coating).\\
In order to extrapolate the ITF TF, the measured TF is fitted from 900~Hz to $\sim1.5$~kHz with points where the coherence 
is higher than 90\%. The fit model is $O_j\times S$ with three free parameters: 
the optical gain, the pole frequency and an extra delay. 
The fit is then extrapolated up to 10~kHz.

\subsubsection{TF during VSR2}
The Virgo TF was measured once per week during VSR2, exciting the WE mirror. 
A typical case of the full TF model is shown in red in figure~\ref{fig:VirgoTF}.
The finesse is fitted between 48 and 51 for all measurements.
The distribution of the fitted delays has an average of $-3\,\mathrm{\mu s}$ and a RMS of $0.5\,\mathrm{\mu s}$.
This systematic non-zero delay highlights a possible few \mus\ error in the (WE) mirror actuation TF
and/or the model used in the fit. 
It is well within the systematic errors estimated on both models.\\

The systematic errors on the ITF TF modulus below 900~Hz are dominated by the 
uncertainties on the mirror actuation parameterization, estimated to $3\%$ in modulus.
At higher frequencies, the parameterization uncertainties are $4\%$ in modulus 
and one has to add the errors coming from the finesse estimation uncertainties: 
assuming $3\%$ error on the finesse, it induces $0.5\%$ uncertainty on the TF modulus.
The total error on the TF modulus above 900~Hz is thus below $4.5\%$.

\subsection{Virgo sensitivity}

The sensitivity curve is the noise level of the GW strain signal $h$ as function of frequency.
It is computed from the noise spectral density of the photodiode readout signal $\mathcal{P}_{AC}$, in $\mathrm{W/\sqrt{Hz}}$,
and the modulus of the ITF TF, in $\mathrm{m/W}$, as:
\begin{eqnarray*}
S_h(f) &=& {\mathcal{P}_{AC}}(f) \times \bigg(TF\bigg[\frac{\mathcal{P}_{AC}}{\Delta_L} \bigg] \bigg)^{-1} \frac{1}{L}
\end{eqnarray*}
where $L$ is the arm cavity length of $3000\,$m for Virgo. 
\begin{figure}[tbh]
\begin{center}
	\includegraphics[width=0.6\linewidth]{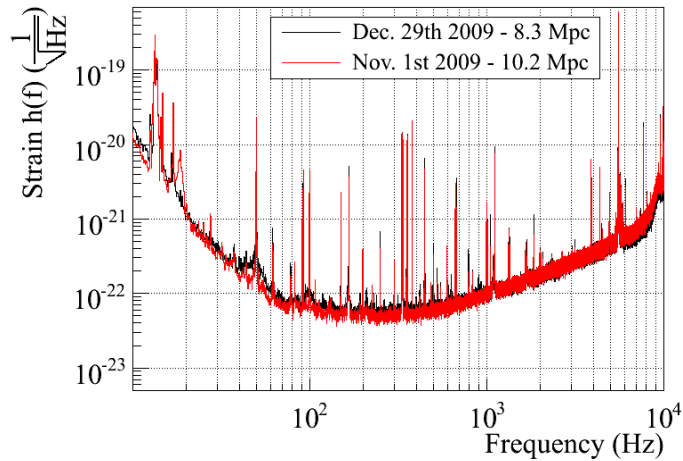}
	\caption{A typical and one of the best Virgo sensitivity curves measured in the range 10~Hz to 10~kHz, 
	given with their associated detection range.
	The 50~Hz line and its harmonics are visible.
	The sets of permanent calibration lines used for the $h(t)$-reconstruction can be seen around 15~Hz, 95~Hz and 355~Hz.
	Some other lines are used to control the Virgo lock (i.e. 379 Hz).
	The other lines are due to environmental noise.
	}
	\label{fig:VirgoSensitivity}
\end{center}
\end{figure}

\subsubsection{Sensitivity during VSR2}

Virgo sensitivity curves measured during VSR2 are shown in figure~\ref{fig:VirgoSensitivity}.
The uncertainties on the sensitivity are directly related to the errors on the Virgo TF modulus,
of the order of 3\% below 900~Hz and up to 4.5\% above.\\

The {\it detection range} $D$ can be extracted as a figure of merit from the ITF sensitivity curve.
It is defined~\cite{bib:DetectionRange} as the distance to which a coalescence of two compact objects of masses $m_1$ and $m_2$
is visible with a signal-to-noise ratio of~8, averaged on the source orientation and direction in the sky:
\begin{eqnarray*}
\frac{D}{1\ \mathrm{Mpc}} &=& \frac{0.25}{2.26} \sqrt{\int_{f_{min}}^{f_{max}}
                               \frac{|\tilde{h}(f)|^2}{S_h(f)}\ \mathrm{d}f}
\end{eqnarray*}
where $S_h(f) $ is the properly calibrated noise one-sided power spectral density
and $\tilde{h}(f)$ is the Newtonian approximation of the signal spectral density for a source at a distance of 1~Mpc:
\begin{eqnarray*} 
|\tilde{h}(f)|^2 = \frac{5}{4}\ \frac{\pi}{6}\ \frac{G_N^{5/3}}{c^3} \frac{\mu M^{2/3}}{d^2}\ (\pi f)^{-7/3}
\end{eqnarray*}
with $ \mu = \frac{m_1 m_2}{m_1+m_2}$, $M = m_1+m_2$, $d = 1\,\mathrm{Mpc}$,
$c$ is the speed of light and $G_N$ the gravitional constant.
The value of $f_{max}$ is set to the frequency at the innermost stable circular orbit,$f_{ISCO}$, defined as:
\begin{eqnarray*}
f_{ISCO} &=& \frac{c^3}{6\sqrt{6}\pi G_N M}
\end{eqnarray*}
In the following, $f_{min}$ is set to 10~Hz.\\

\begin{figure}[tbh]
\begin{center}
	\includegraphics[width=0.8\linewidth]{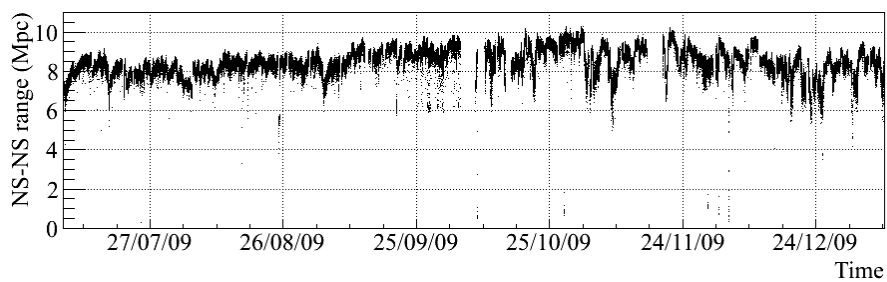}
	\caption{
	Virgo BNS detection range during VSR2 \SciMode\ data, estimated every minute.
	}
	\label{fig:RangeVsTime}
\end{center}
\end{figure}

\begin{figure}[tbh]
\begin{center}
	\includegraphics[width=0.5\linewidth]{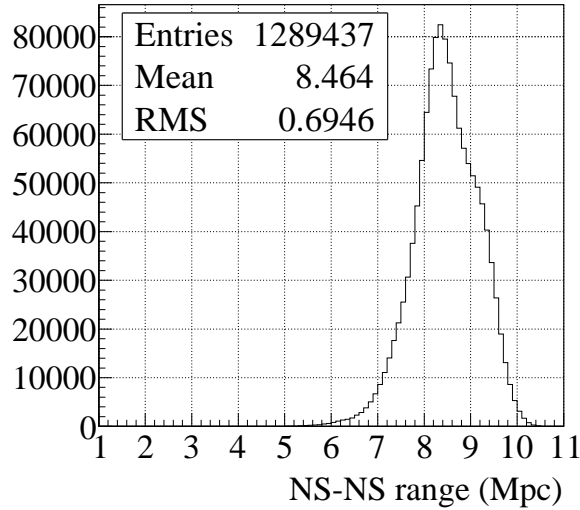}
	\caption{
	Distribution of Virgo BNS detection range estimated every 10~s during VSR2 \SciMode\ data.
	}
	\label{fig:RangeDistri}
\end{center}
\end{figure}

The binary neutron star (BNS) detection range (for a $1.4\,\mathrm{M}_{\odot}\times1.4\,\mathrm{M}_{\odot}$ system) 
is used in figure~\ref{fig:RangeVsTime} to show the Virgo sensitivity over time  during VSR2 (149.3~days of \SciMode\ data). 
Its distribution is shown in figure~\ref{fig:RangeDistri}: its average is $8.5$\,Mpc and its RMS $0.7$\,Mpc.
The BNS detection ranges were between 7.1~Mpc and 9.9~Mpc for 95\% of the \SciMode\ time.
A typical and one of the best sensitivity curves are shown in figure~\ref{fig:VirgoSensitivity}.
The BNS detection range is mainly sensitive to the noise level between $\sim50$~Hz and $\sim500$\,Hz, 
where the systematic errors, coming from $S_h(f)$, are estimated to 4\%.\\

\begin{figure}[tbh]
\begin{center}
	\includegraphics[width=0.6\linewidth]{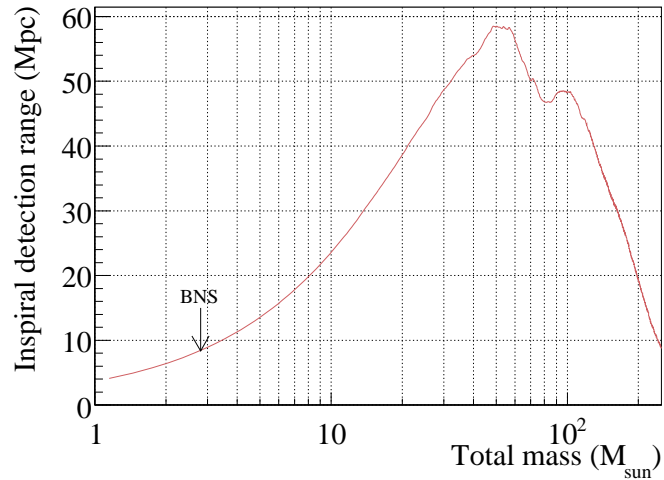}
	\caption{
	Virgo detection range as function of the binary system mass $M$ (with $m_1=m_2=M/2$)
	in the case of a typical VSR2 sensitivity (BNS detection range of 8.3~Mpc).
	}
	\label{fig:RangeVsMass}
\end{center}
\end{figure}

The detection range computed as function of the total mass of the binary system 
is shown figure~\ref{fig:RangeVsMass} for a typical sensitivity of VSR2.
When the component mass is higher, the upper cut-off frequency becomes smaller:
it means that the inspiral range focuses on a narrower low-frequency band of the sensitivity.

\section{Summary}

Calibration procedures have been developed and implemented for the Virgo ITF 
during its first~\cite{bib:CalibVSR1} and second science runs. They are used to estimate the absolute scale
of the displacements to which the ITF is sensitive, down to $\sim 10^{-19}$\,m.\\

The methods used to measure the mirror actuation response, the ITF output power sensing and absolute timing 
have been shown along with their performances.
The stability of the parameters during VSR2 has been checked.
When available, independent measurements of the parameters are in good agreement.
They are used to estimate the systematic errors which are dominant compared to the statistical uncertainties.
Typical results obtained during VSR2 have been shown.
An important parameterization is the mirror actuation response below 900~Hz, 
known within $3\%$ in modulus and $14$\,mrad/$10$\,\mus\ in phase.
The {\it dark fringe} power sensing has been measured with a timing precision of 4\,\mus.\\

The frequency-domain sensitivity estimation, $S_h(f)$, has been described.
Systematic errors have been estimated to be of the order of 3\% below 900~Hz, 
increasing up to 4.5\% at higher frequencies.
Typical and best sensitivities of Virgo during VSR2 have been shown.
As a figure of merit, the BNS detection range of Virgo during VSR2 (149~days of \SciMode\ data)
had an average value of $8.3$\,Mpc, with a RMS of $0.7$\,Mpc.



\end{document}